\documentclass[english,aip,letter,superscriptaddress,twocolumn,floatfix,jcp,nourl,nofootinbib]{revtex4}
\usepackage[utf8]{inputenc}
\usepackage[T1]{fontenc}
\usepackage{upgreek}
\usepackage{graphicx}
\usepackage{ae,aecompl}
\usepackage{color}
\usepackage{pstricks,pst-node}
\usepackage{psfrag}
\usepackage{verbatim}
\usepackage{bm}
\usepackage{pdflscape}
\usepackage{rotating}
\usepackage{stmaryrd}
\usepackage{amsfonts}
\usepackage{amsmath}
\usepackage{amssymb}
\usepackage{bbold}
\usepackage{endnotes}
\usepackage{ifthen}
\usepackage{bm}
\usepackage{setspace}
\usepackage{endnotes}
\usepackage{fancyhdr}
\usepackage{fancybox}
\usepackage{makeidx}
\usepackage{framed}
\usepackage{listings}
\usepackage{alphalph}
\definecolor{shadecolor}{rgb}{0,.7,0}
\definecolor{shadecolor}{gray}{0.7}
\definecolor{shadecolor}{gray}{0.95}
\definecolor{ogreen}{rgb}{0,0.8,0}
\definecolor{magenta}{rgb}{1,0,1}
\definecolor{brown}{rgb}{0.7,0.4,0.2}
\definecolor{shadecolor}{gray}{0.9}

\newcommand{\del}[1]{{\color{red}#1}}
\newcommand{\new}[1]{{\color{blue}#1}}

\newcommand{\Note}[1]{{\bf \color{red}#1}}

\newcommand{\esc}{\!\cdot\!}

\RequirePackage{dsfont}

\usepackage{xr-hyper}
\usepackage{hyperref}
\graphicspath{{../00-Figuras-new/}}
\begin{document}

\title{Internal dissipation in the tennis racket effect}

\author{J.A. de la Torre and Pep Espa\~nol}
\affiliation{ Dept.   F\'{\i}sica Fundamental, Universidad Nacional
  de Educaci\'on a Distancia, Madrid, Spain}
\date{20th March 2021}
\date{\today}
\begin{abstract}
  The phenomenon known as the tennis  racket effect is observed when a
  rigid  body experiences  unstable rotation  around its  intermediate
  axis. In  free space,  this leads to  the Dzhanibekov  effect, where
  triaxial objects  like a  spinning wing  bolt may  continuously flip
  their rotational axis.  Over time, however, dissipation ensures that
  a torque free spinning body  will eventually rotate around its major
  axis, in  a process called precession  relaxation, which counteracts
  the  tennis  racket effect.   Euler's  equations  for a  rigid  body
  effectively describe  the tennis  racket effect, but  cannot account
  for  the precession  relaxation  effect.  A  recent  theory has  put
  forward  a   generalization  of  Euler's  equations   that  includes
  dissipation  in  a  thermodynamically consistent  way.   The  theory
  displays  two dissipative  mechanisms:  orientational diffusion  and
  viscoelasticity.  Here we show  that orientational diffusion, rather
  than  viscoelasticity, primarily  drives  precession relaxation  and
  effectively suppresses the tennis racket effect.
 \end{abstract}
\maketitle

\section{Introduction}
\begin{figure}[t]
  \includegraphics[width=0.48\textwidth]{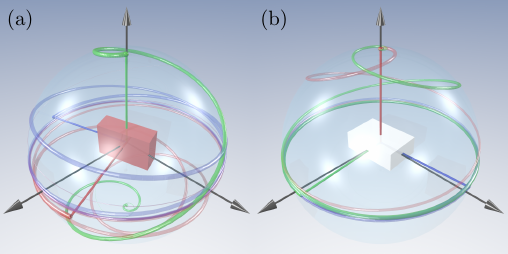}

  \caption{ A triaxial body is set in motion with the angular momentum
    in  the $z$  direction and  with the  intermediate principal  axis
    (green) initially also in the  $z$ direction.  The coloured traces
    are the  trajectories of the  three unit principal  vectors.  {\bf
      (a)}  A \textit{rigid}  body shows  a periodic  flipping of  the
    intermediate axis known as the Dzhanibekov effect, as predicted by
    Euler's  equations.  The  intermediate  principal  axis vector  is
    initially upwards and after some  rotations of the body, it points
    downwards.   The flipping  of  the axis  repeats itself  endlessly
    although  only one  flip  is shown  in the  figure.   {\bf (b)}  A
    \textit{quasi-rigid}  body  shows precession  relaxation  instead,
    where the  body ends up  spinning along  the major axis  (red), as
    predicted by the dissipative Euler's  equations.  A movie of these
    phenomena is presented in the Supplementary Material.}
  \label{Fig:DE}
\end{figure}

The dynamics of  a rigid body in motion are  elegantly captured by the
classical  Euler's  equations  \cite{goldsteinClassicalMechanics1983},
which  illuminate the  complex and  fascinating behaviors  of triaxial
bodies,  especially when  rotating about  their
intermediate  axis.   The  motion  around  the  intermediate  axis  is
unstable
\cite{poinsot2022outlines,landauMechanicsThirdEdition1960,Ashbaugh1991}
and  leads to  the ``tennis  racket  effect'' where  a tennis  racket,
thrown  from its  handle with  a  spin around  its intermediate  axis,
exhibits an unexpected flipping motion in mid-air, as it traverses the
path  back to  the hand  of  the tennis  player.  Another  spectacular
demonstration  of  this  effect  was  reported  by  Russian  cosmonaut
Dzhanibekov  who observed  in 1985  how a  suddently released  wing nut
spins rapidly around its central axis and keeps flipping its orientation.
Striking  videos on  the  internet  show the  effect  in zero  gravity
environments  \cite{veritasium, physicsunsimplified}.       In      Fig.
\ref{Fig:DE} (a) we show a  pictorial representation of the Dzhanibekov
effect.
There   is   recent   interest   in  fully   describing   the   tennis
racket/Dzhanibekov  effect (DE)  incorporating theoretical  analysis
\cite{vandamme2017,mardesic2020a},         numerical         solutions
\cite{Murakami2016}     applicable      to     spacecraft     dynamics
\cite{Trivailo2019}, molecular dynamics simulations \cite{lun-fu2022},
and through experiments using mobile phones \cite{wheatland2021}.

However, real  bodies are not  completely rigid and  Euler's equations
are  just an  approximation.  The  lack  of rigidity  arises not  only
because  of  the elastic  response  of  the  body  but also  from  the
intrinsic thermal  fluctuations experienced by the  constituent atoms.
These  intrinsic  fluctuations  lead  to  dissipative  processes  that
gradually convert rotational kinetic energy into thermal energy.  As a
result, the system naturally  progresses towards a state characterized
by  minimal kinetic  energy  which corresponds  to  the body  spinning
around         its        major         axis        of         inertia
\cite{landauMechanicsThirdEdition1960,Lamy1972}.   This phenomenon  is
referred to as nutational damping  or precession relaxation, and it is
shown in Fig.  \ref{Fig:DE} (b). Had Dzhanibekov had observed his wing
nut for  a sufficiently long period,  he would have found  it spinning
around the major axis.   In other words, \textit{precession relaxation
  kills the Dzhanibekov effect in the long run}.

Precession relaxation  explains why roughly  98\% of asteroids  in the
Light  Curve  Database LCDB  \cite{Warner2009}  are  in pure  rotation
\cite{Kwiecinski2020}.  It  is also responsible for  some catastrophic
design problems in artificial satellites  in early times of spacecraft
history  \cite{Efroimsky2001}.   Precession  relaxation  is  currently
attributed  to the  dissipation caused  by inelastic  relaxation, i.e.
viscoelasticity
\cite{Efroimsky2001,Prendergast1958,Burns1973,efroimsky2000,Efroimsky2002,Molina2003,Sharma2005,Lazarian2007,Breiter2012,Efroimsky2015,sharmaShapesDynamicsGranular2017,frouard2018,Kwiecinski2020}.
Inelastic relaxation  results from  the alternating stresses  inside a
wobbling  body, caused  by transversal  and centripetal  acceleration,
leading  to  deformation  and  energy dissipation.   For  a  solid  of
revolution, the  angle of precession  between the principal  axis with
largest inertia  moment and the  conserved angular momentum  vector is
univocally    related    to     the    rotational    kinetic    energy
\cite{Efroimsky2001}   and,  therefore,   by   computing  the   energy
dissipated one can  infer the rate of change of  the precession angle.
Recent     attempts     \cite{frouard2018,Kwiecinski2020}     evaluate
approximately the  dissipated energy  by solving the  continuum stress
${\bf P}$ and strain  $\dot{\boldsymbol{\epsilon}}$ fields of a linear
viscoelastic  model for  an  ellipsoid under  the non-inertial  forces
appearing in the  principal axis frame.  The power  dissipated is then
identified         with         the         entropy         production
$T\sigma=                \boldsymbol{\Pi}:\dot{\boldsymbol{\epsilon}}$
\cite{deGroot1964}  and the  relaxation rate  is estimated.   However,
using \textit{dissipative} continuum field theories to describe a body
obeying the  \textit{reversible} Euler's  equations specific  to rigid
bodies presents a certain inconsistency, justified from practical necessity only.

In  this  Letter, we  present  a  significantly distinct  approach  to
addressing the issue of  precession relaxation.  Our approach modifies
Euler's  equations  using  non-equilibrium  statistical  mechanics  to
incorporate dissipation in a thermodynamically consistent manner.  Two
distinct   dissipative   mechanisms  are   identified:   orientational
diffusion and viscoelasticity.  Orientational  diffusion refers to the
microscopic   process  through   which  thermal   fluctuations  induce
alterations in the orientation of a  body, even when the body has zero
angular momentum  and does  not spin.  This  effect is  appreciable in
complex molecules  in a  vacuum, where thermal  fluctuations gradually
reshape  the  molecule, ultimately  resulting  in  alterations to  its
overall orientation \cite{SaportaKatz2019, Peng2021}.  The exploration
of all  possible orientations is imperceptible  for macroscopic bodies
due to the  exceedingly long time scales involved.   However, we argue
that  orientational diffusion  is the  fundamental process  underlying
precession relaxation  in macroscopic  bodies. The  second dissipative
mechanism in  this theory is  due to dilational  friction, responsible
for the damping of elastic oscillations of the body.  We refer to this
second  dissipative  mechanism  as   viscoelasticity.   We  show  that
viscoelasticity plays  no role in precession  relaxation. An effective
way to  support our  claim is  by examining  the impact  of precession
relaxation on the DE.

\textit{Dissipative Euler's equations.-} The equations of motion for a
free quasi-rigid body that generalize Euler's equations to account for
internal   thermal   noise   and   dissipation   are   formulated   in
\cite{espanol2024}.   In  this  theory,  the  state  of  the  body  is
described with the  orientation and the shape of  the body, determined
by the eigenvectors ${\bf  e}_\alpha$ and eigenvalues ${\bf M}_\alpha$
of the gyration tensor, respectively.   The gyration tensor is defined
microscopically                                                     as
$   {\bf   G}=\frac{1}{4}\sum_{i}{\bf  r}_i\otimes{\bf   r}_i$   where
${\bf r}_i$ is  the position of the $i$-th particle  of the body.  The
inertia   tensor    is   related    to   the   gyration    tensor   as
${\bf  I}=4\left({\rm  Tr}[{\bf G}]\mathbb{1}-{\bf  G}\right)$,  where
${\rm Tr}[\cdots]$ denotes the trace of the matrix and $\mathbb{1}$ is
the identity  matrix.  For  macroscopic bodies,  the following  set of
ordinary  differential equations  (ODE)  governs the  dynamics of  the
orientation $\boldsymbol{\Lambda}$ of the body and the central moments
${\bf M}$
\begin{align}
  \label{ODE}
   \frac{d\boldsymbol{\Lambda}}{dt}
    & = {\bf B}\esc\left[
\boldsymbol{\Omega}
      -      
\boldsymbol{\cal D}\esc(\boldsymbol{\Omega}\times{\bf S})\right],
      \nonumber\\
   \frac{d{\bf M}}{dt}&=\boldsymbol{\Pi},
        &&
 \frac{d\boldsymbol{\Pi}}{dt}  ={\boldsymbol{\cal K}} -\boldsymbol{\Gamma}\esc\boldsymbol{\Pi}.
\end{align}
The  orientation  $\boldsymbol{\Lambda}$  parameterizes  the  rotation
matrix $\boldsymbol{\cal R}=e^{[-\boldsymbol{\Lambda}]_\times}$, where
$[\cdots]_\times$   is  the   cross   product   matrix.   The   matrix
$\boldsymbol{\cal    R}$   contains    as   rows    the   eigenvectors
${\bf e}_\alpha$  of the gyration tensor  and, therefore, diagonalizes
it                             according                            to
$ \mathbb{G}= \boldsymbol{\cal  R}\esc{\bf G}\esc\boldsymbol{\cal R}^T
$  where the  diagonal  matrix $\mathbb{G}$  has  the central  moments
${\bf M}$ in the diagonal.  The dilational momentum $\boldsymbol{\Pi}$
is defined  as the time derivative  of the central moments  ${\bf M}$.
In    (\ref{ODE})     the    spin     velocity    is     defined    as
$\boldsymbol{\Omega}={\bf I}^{-1}\esc{\bf  S}$ where ${\bf S}$  is the
conserved  angular momentum  of  the body.  The \textit{dynamic}  spin
velocity  $\boldsymbol{\Omega}$  should   be  distinguished  from  the
\textit{kinematic}  angular  velocity   $\boldsymbol{\omega}$  of  the
principal axis  system which is defined  in the usual way  in terms of
the                           rotation                          matrix
$        [\boldsymbol{\omega}]_\times\equiv        -{{\boldsymbol{\cal
      R}}}^{T}\esc\frac{d{{\boldsymbol{\cal                 R}}}}{dt}$
\cite{v.i.arnoldMathematicalMethodsClassical1989,Diaz2019}.        The
angular velocity is related to  the time derivative of the orientation
according                                                           to
$\frac{d{{\boldsymbol{\Lambda}}}}{dt}       ={\bf       B}       \cdot
{\boldsymbol{\omega}}$,    where    the    kinematic    operator    is
\cite{Diaz2019,espanol2024}
\begin{align}
  \label{eq:53}
  {\bf B}
  &=\mathbb{1}-\frac{{\Lambda}}{2} [{\bf n}]_\times+\left(1-\frac{{\Lambda}}{2}\frac{\sin{\Lambda}}{(1-\cos{\Lambda})}\right)[{\bf n}]_\times[{\bf n}]_\times
\end{align}
where               $\Lambda=|\boldsymbol{\Lambda}|$               and
${\bf n}=\boldsymbol{\Lambda}/\Lambda$.  The  angular diffusion tensor
is                              defined                             as
$  \boldsymbol{\cal  D}=  \boldsymbol{\cal  R}^T\esc  \boldsymbol{\cal
  D}_0\esc \boldsymbol{\cal R} $ where the angular diffusion tensor in
the       principal      axis       frame      has       the      form
$  \boldsymbol{\cal D}_0({\bf  M},{\cal E})={\rm  Diag}[d_1,d_2,d_3]$,
with  $d_\alpha>0$.   The  dilational  friction  matrix  is  given  by
$  \boldsymbol{\Gamma}={\rm   Diag}[\gamma_1,\gamma_2,\gamma_3]$  with
$\gamma_\alpha>0$.  Finally,  the dilational  force has  the following
components
\begin{align}
  \label{eq:15}
      {\boldsymbol{\cal K}}^\alpha
  &={\bf M}_{\alpha}\left( \frac{1}{2}(\boldsymbol{\nu}^{\alpha})^2
    +2\left(    \boldsymbol{\Omega}_p^T\esc \boldsymbol{\Omega}_p -( \boldsymbol{\Omega}_p^\alpha )^2\right)
    +\boldsymbol{\sigma}^\alpha\right )
\end{align}
Here   the   spin   velocity   in  the   principal   axis   frame   is
${\boldsymbol{\Omega}_p}\equiv
e^{-[\boldsymbol{\Lambda}]_\times}\cdot\boldsymbol{\Omega}$,       the
dilational velocity is defined as  the ratio of dilational momentum to
central                                                        moments
$   \boldsymbol{\nu}_\alpha=\frac{\boldsymbol{\Pi}_\alpha}{M_\alpha}$,
and the elastic acceleration is
\begin{align}
\boldsymbol{\sigma}=\boldsymbol{\Sigma}^{-1}\cdot\left({\bf M}-{\bf M}^{\rm rest}\right)  
\end{align}
where  $\boldsymbol{\Sigma}$ is  the equilibrium  covariance of  central
moments, which  plays the role  of a  matrix of elastic  constants. In
general,  the   matrix  $\boldsymbol{\Sigma}$  has  all   the  entries
different from zero, because compressions of the body in one direction
may affect  the expansion  in others.   However, and  for the  sake of
simplicity, we will consider a model of elasticity in which the matrix
is                                                           diagonal,
$\boldsymbol{\Sigma}={\rm   Diag}[\Sigma_1,\Sigma_2,\Sigma_2]$.    The
dilational force $\boldsymbol{\cal K}$  has a centrifugal contribution
depending  on ${\boldsymbol{\Omega}_p}$  and  an elastic  contribution
that tries  to restore the  value of the  central moments to  its rest
value ${\bf  M}^{\rm rest}$. The  motion of central moments  is damped
with  the friction  force $-\boldsymbol{\Gamma}\cdot\boldsymbol{\Pi}$.
The central moments therefore evolve  in a damped oscillatory way that
we refer to as the  \textit{viscoelastic} behaviour of the present model.  As
the    inertia    tensor    depends   on    both    the    orientation
$\boldsymbol{\Lambda}$ and the central moments ${\bf M}$, and the spin
velocity $\boldsymbol{\Omega}={\bf I}^{-1}\cdot{\bf S}$ appears in the
dynamics of  both variables, the  dynamics of orientation  and central
moment are fully coupled.

\newpage\textit{Euler's equations.-}  Euler's equations  are obtained under two
assumptions \cite{espanol2024}.  The first  assumption ${\cal H}_1$ is
that    the   angular    velocity   and    spin   velocity    coincide
$\boldsymbol{\omega}=\boldsymbol{\Omega}$.  This  can  be  written  as
$ \frac{d\boldsymbol{\Lambda}}{dt} =  {\bf B}\cdot \boldsymbol{\Omega}
$, which  is a tiny  bit of the set  of ODEs (\ref{ODE}). By using
the definition of spin angular velocity and the diagonalization of inertia tensor gives
\begin{align}
    \label{Euler}
  \frac{d\boldsymbol{\Lambda}}{dt}
  & = {\bf B}\esc
    e^{[\boldsymbol{\Lambda}]_\times}\esc\mathbb{I}^{-1}\esc e^{-[\boldsymbol{\Lambda}]_\times}\esc{\bf S}
\end{align}
where  $\mathbb{I}$ is  the diagonalized  inertia tensor.   The second
assumption ${\cal H}_2$  is that the central moments do  not change in
time  ${\bf  M}(t)={\bf  M}^{\rm  rest}$,  and  $\mathbb{I}$  is  time
independent.   In  this case,  the  ODE  (\ref{Euler}) is  closed  for
$\boldsymbol{\Lambda}$, which is  entirely  equivalent  to  Euler's
equations but  provides directly  the orientation  of the  rigid body.
Therefore,  (\ref{ODE})  generalizes  Euler's equations  by  including
dissipation  in a  thermodynamically consistent  way. To  the authors'
knowledge,      the     \textit{orientational      diffusion}     term
$\boldsymbol{\cal   D}\cdot(\boldsymbol{\Omega}\times{\bf    S})$   in
(\ref{ODE}) is new.

\textit{Thermodynamic consistency.}  The  set of equations (\ref{ODE})
comply with  the Second Law of  thermodynamics. The entropy of  a free
macroscopic   body   at   the   present  level   of   description   is
\begin{align}\label{entropy}
S_B     =S^{\rm MT}({\cal E})  -({{\bf     M}}-{\bf      M}^{\rm
  rest})^T\esc\frac{\boldsymbol{\Sigma}^{-1}}{2T^{\rm   MT}}\esc({{\bf
    M}}-{\bf M}^{\rm rest}),  
\end{align}
where $S^{\rm  MT}({\cal E})$ is the  usual macroscopic thermodynamics
entropy   of  the   body,   that  depends   on   the  thermal   energy
${\cal  E}=  E- K^{\rm  rot}-K^{\rm  dil}$,  which  is the  result  of
substracting    the   ``organized    forms    of   kinetic    energy''
$K^{\rm rot},K^{\rm  dil}$ from the  total conserved energy  $E$.  The
rotational    kinetic     energy    has    the     usual    expression
$ K^{\rm rot} =\frac{1}{2}{\bf  S}^T\cdot{\bf I}^{-1}\cdot{\bf S}$ and
the dilational kinetic energy associated  with changes in the shape of
the                               body                              is
$                              K^{\rm                             dil}
=\frac{1}{2}{\boldsymbol{\Pi}}^T\cdot\mathbb{G}^{-1}\cdot{\boldsymbol{\Pi}}
= \sum_\alpha\frac{\boldsymbol{\Pi}_\alpha^2}{2{\bf  M}_\alpha}$.  The
thermodynamic   temperature  is   given   by   the  usual   definition
$  \frac{1}{T^{\rm MT}}=  \frac{\partial S_B^{\rm  MT}}{\partial {\cal
    E}}$.  The time derivative of the entropy can be computed from the
dynamics (\ref{ODE}) and the chain rule leading to
\begin{align}
  \label{eq:313}
T^{\rm MT}\frac{dS_B}{dt}&=    \left(\boldsymbol{\Omega}\times{\bf S}\right)^T\esc\boldsymbol{\cal D}\esc    \left(\boldsymbol{\Omega}\times{\bf S}\right)
+     \boldsymbol{\nu}^T\esc\boldsymbol{\Gamma}\esc\boldsymbol{\nu}\ge0
\end{align}
This  time derivative  is always  positive,  as a  consequence of  the
positive      character      of     the      dissipative      matrices
$\boldsymbol{\cal  D},\boldsymbol{\Gamma}$.   Therefore,  the  entropy
plays the role of a Lyapunov  function for the ODEs (\ref{ODE}), which
comply with the  Second Law.  The system reaches  an equilibrium state
at long times where the  entropy is maximal. From (\ref{eq:313}), this
corresponds           to          the           conditions          i)
$\boldsymbol{\Omega}^{\rm     eq}\times{\bf      S}=0$     and     ii)
$\boldsymbol{\nu}^{\rm eq}=0$.  The first condition i) states that the
equilibrium  value of  the spin  velocity  is paralel  to the  angular
momentum vector, which  can only occur if the body  aligns to have the
major  principal axis  in  the  direction of  ${\bf  S}$.  The  second
condition   ii)   implies   that   the   central   moments   reach   a
time-independent equilibrium value.

We claim that the actual  responsible for precession relaxation is not
viscoelasticity  but  rather  orientational diffusion.   This  may  be
suggested  by  the  form  of  the  entropy  production  (\ref{eq:313})
displaying  the two  mechanisms,  and the  fact  that the  equilibrium
condition  $\boldsymbol{\Omega}^{\rm  eq}\times{\bf  S}=0$ can  only  be
achieved if  $\boldsymbol{\cal D}_0\neq0$.   However, this  is further
substantiated numerically by showing  that switching off orientational
diffusion but switching  on dilational friction does not  kill the DE,
while doing otherwise (orientational diffusion on, dilational friction
off) leads to precession relaxation, and cessation of the DE.

\textit{Set  up.-} The  triaxial  body has  dimensions $(a,b,c)$  with
$b=2a,  c=4a$.   The model  (\ref{ODE})  contains  a large  number  of
parameters.    To  simplify   our   analysis   we  set   $d_\alpha=d$,
$\gamma_\alpha=\gamma$,   $\Sigma_\alpha=\Sigma$.     Typical   values
include  \(d=0.1\), \(\gamma=0.05\),  and  \(\Sigma=0.1\).  These  are
selected for  numerical convenience since realistic  values can result
in vastly separated time scales.  We choose units such that total mass
$M=1$, spin  velocity $\Omega=1$ and $M_3^{\rm  rest}=1$.  The angular
momentum  is chosen  in the  $z$ direction,  ${\bf S}=(0,0,S)$  which,
together with total energy $E$ fix the equilibrium state.  We consider
the evolution  of a  body that  is initially set  in motion  through a
rotation  around the  intermediate axis.   In this  way, we  study the
effect  of  dissipation  on   the  Dzhanibekov  effect.   The  initial
conditions  needed by  the  ODE (\ref{ODE})  that  correspond to  this
situation       are      $       \boldsymbol{\Lambda}(0)=(\pi/2,0,0)$,
${\bf     M}(0)={\bf     M}^{\rm     rest}=\frac{m}{12}(a^2,b^2,c^2)$,
$\boldsymbol{\Pi}(0)=0$.   Once  the  body  is set  into  motion,  the
centrifugal term, proportional  to the square of the  spin velocity in
the dilational  force \(\boldsymbol{\cal K}\)  (\ref{eq:15}), triggers
oscillatory motion in the central  moments. If the dilational friction
coefficient is  non-zero, this motion  will eventually dampen  and the
central moments will  reach equilibrium values that  differ from their
initial rest values due to the influence of centrifugal forces.

\textit{Results.-}  In each  column of  Fig.  \ref{Fig:cases}  we plot
each of the five different configurations considered. The columns show
the  principal  vectors  ${\bf   e}_1$  (red),  ${\bf  e}_2$  (green),
${\bf e}_3$  (blue) plotted both, as  a trajectory in 3D  space and as
their  three  components   $(e^1_\alpha,e^2_\alpha,e^3_\alpha)$  as  a
function of time.  Also shown at  the bottom panels are the rotational
$K^{\rm rot}$  and dilational $K^{\rm  dil}$ kinetic energies  in each
case.  The  column (a)  in Fig.   \ref{Fig:cases} shows  the numerical
solution  of Euler's  equations  (\ref{Euler})  while columns  (b)-(e)
display the solution of  the dissipative Euler's equations (\ref{ODE})
for different  values of the dissipative  coefficients $d,\gamma$.
Euler's equation  in Fig.  \ref{Fig:cases}  (a) exhibits the  DE, most
clearly seen  through the time  evolution of the  (green) intermediate
eigenvector ${\bf  e}_2$, that  keeps flipping its  direction. Euler's
equations conserve both rotational and dilational kinetic energies, as
shown at  the bottom panels  of column (a). Fig.   \ref{Fig:cases} (b)
displays  the   solution  of  (\ref{ODE})  when   $d=0,\gamma=0$  that
corresponds to a purely reversible  dynamics.  The result is different
from  Euler's solution  in  Fig. \ref{Fig:cases}  (a)  because of  the
coupling of the  dynamics of the orientation and  of the central
moments.  The intermediate  axis shows the flipping  effect typical of
the  DE.    The  dilational  and  rotational   kinetic  energies  show
oscillations due  to the  undamped oscillatory  motion of  the central
moments.
\begin{widetext}
\newpage\begin{figure*}[h]
  \includegraphics[width=0.94\textwidth]{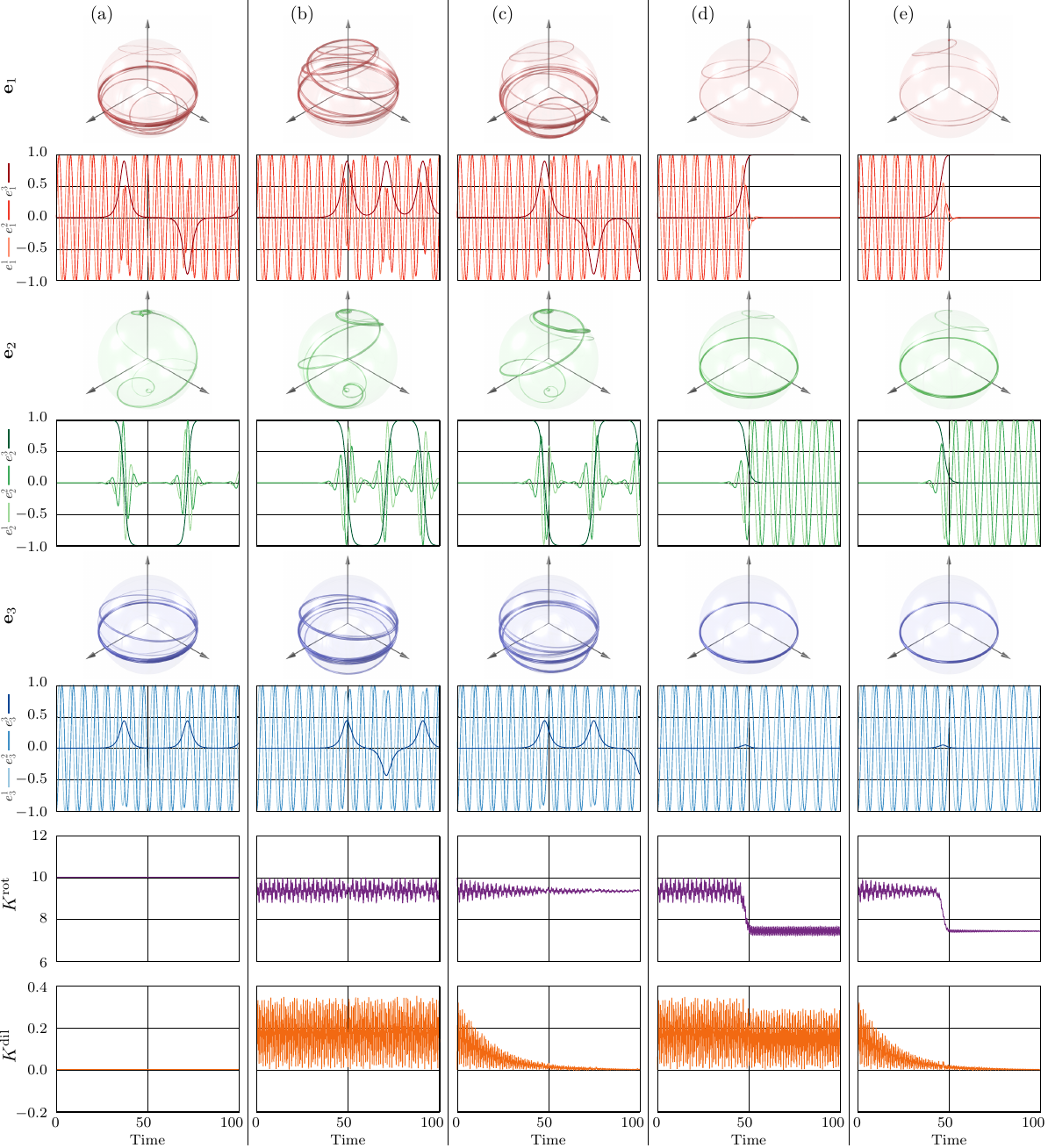}
  \caption{  {\bf (a)}  Solution  of  Euler's equations  (\ref{Euler})
    describing the DE.  The  solution of dissipative Euler's equations
    (\ref{ODE}) for  different values  of the  parameters is  shown in
    {\bf  (b)}:  $d=0,\gamma=0$,  {\bf (c)}:  $d=0,\gamma=0.05$,  {\bf
      (d)}: $d=0.1,\gamma=0.0$,  {\bf (e)}:  $d=0.1,\gamma=0.05$.  Red
    panels show both the trajectory of ${\bf e}_1(t)$ in space and its
    three   components   $(e^1_1,e^2_1,e^3_1)$   as  a   function   of
    time. Green panels are for ${\bf e}_2(t)$, and blue panels are for
    ${\bf e}_3(t)$.  Bottom panels  show the rotational  $K^{\rm rot}$
    and dilational $K^{\rm dil}$ kinetic energies as a function of time
    in each case. }

\label{Fig:cases}
\end{figure*}
\end{widetext}
Fig.  \ref{Fig:cases}  (c) corresponds  to no  orientational diffusion
$d=0$ but  non-zero viscoelasticity $\gamma=0.05$.  In  this case, the
DE is  still present and  visible.  The dilational friction  damps the
oscillations of central  moments, that attain an  equilibrium value in
the long run  as shown in the dilational kinetic  energy at the bottom
panel of column (c).  The corresponding rotational kinetic energy also
shows some damping corresponding  to equilibration of central moments,
but oscillations are  still present due to the  undamped precession. A
very  different dynamical  situation emerges  when we  have rotational
diffusion  $d=0.1$ and  no viscoelasticity  $\gamma=0$. In  this case,
shown in column (d), the  (red) principal vector ${\bf e}_1$ initially
rotates in a circumference but suddently adopts the vertical position,
while  the initially  vertical (green)  intermediate principal  vector
${\bf  e}_2$  ends  up  moving   in  circles.   The  body  experiences
precession relaxation instead of the  DE shown in previous cases. This
is clearly manifest in the decrease of rotational kinetic energy shown
at the bottom of column (d) when the re-orientation of the body occurs
and the  rotation axis changes  from ${\bf  e}_2$ to ${\bf  e}_1$.  As
there is  no dilational  friction in this  case, central  moments keep
oscillating,  as reflected  in the  dilational and  rotational kinetic
energies  at   the  bottom  panels   of  column  (d).    Finally,  for
completeness,  Fig.  \ref{Fig:cases}  (e) illustrates  the case  where
both  dissipative  mechanisms  are  active,  with  parameters  set  at
\(d=0.1\)  and  \(\gamma=0.05\),  which  is also  the  case  shown  in
Fig. \ref{Fig:DE}  (b).  In  this case, precession  relaxation occurs,
and the  central moments are  damped by dilational friction.   This is
reflected in  the dilational  kinetic $K^{\rm  dil}$ that  vanishes at
long  times,  while $K^{\rm  rot}$  experiences  a decrease  when  the
re-orientation  of the  body  occurs. At  long  times, the  rotational
kinetic energy is constant due to the constancy of both, spin velocity
and the principal moments.

In summary,  if there  is no orientational  diffusion, $d=0$  the body
displays  the  DE, while  if  $d\neq0$  the body  displays  precession
relaxation,  irrespective   of  the  values  of   dilational  friction
$\gamma$.   A video  illustrating the  phenomenology described  in Fig.
\ref{Fig:cases} is provided as Suplemental Material.

\bigskip  \textit{Conclusions.-}  The  dissipative  Euler's  equations
(\ref{ODE}) governs the  dynamics of  the orientation  and shape  of
quasi-rigid  bodies   described  with  the  gyration   tensor.   These
equations, derived from non-equilibrium statistical mechanics, capture
both spatial orientation changes and  the damped oscillatory motion of
the  body's  central  moments.  The  equations  are  thermodynamically
consistent  and  respect  the  Second Law.   Two  physically  distinct
mechanisms  describe the  internal dissipation  that leads  to entropy
production:     orientational    diffusion     and    viscoelasticity.
Orientational diffusion is controlled by a new term to be added to the
Euler's equations for a rigid body,  and whose effect is to reduce the
rotational  kinetic  energy  of  the body.   In  the  present  theory,
viscoelasticity  is   described  by   the  elastic  dynamics   of  the
eigenvalues  of the  gyration tensor  which  is damped  with a  simple
dilational friction mechanism.

A body spinning around the intermediate  axis and prone to display the
DE is  a good scenario  on which to test  the effect of  the different
dissipative mechanisms.  We  have observed that the  DE dissapears due
to  precession  relaxation  only   when  the  orientational  diffusion
coefficient is non-zero,  irrespective of the value  of the dilational
friction.   This  shows  that  the  precession  relaxation  phenomenon
predicted from (\ref{ODE}) stems   from
orientational diffusion, as opposed to viscoelastic dissipation.

The current view describing the alignment process of celestial bodies,
and estimating  the corresponding  relaxation times,  is based  on the
idea  that  inelastic  relaxation   arises  from  alternating  elastic
stresses  generated inside  a  wobbling body  by  the transversal  and
centripetal acceleration of its parts.  For a viscoelastic solid, this
will  dissipate   rotation  kinetic  energy,  leading   to  precession
relaxation.   It  is   important  to  recognize  that   this  type  of
viscoelasticity  based  on the  stress  and  strain tensor  fields  is
conceptually, and  quantitatively, different from  the viscoelasticity
in our  model.  These  two concepts of  viscoelasticity belong  to two
different  coarse-grained  levels  of description,  a  detailed  level
characterized  with  continuum  fields,   and the present  coarser  level
described with  the gyration  tensor.  We emphasize  that there  is no
such a thing as \textit{the} entropy  of a system.  Rather, each level
of  description has  its  own  entropy function,  as  attested by  the
distinct functional dependence  of the entropy on  the state variables
of  the   corresponding  level   of  description.   This   is  clearly
illustrated in our  case: the entropy (\ref{entropy})  of the gyration
tensor level  of description is  \textit{not} given by the entropy  of the
macroscopic  level  of  description,  as they  differ  by  an  elastic
contribution.    Therefore,   entropy   and   its   production,   i.e.
dissipation, are \textit{relative} to  the chosen level of description
-- and so it is viscoelasticity.

It is  possible to extract  from (\ref{ODE}) the  dynamics of
the precession angle between the  major principal axis and the angular
momentum.  This angle is fully determined by the moments of inertia of
the  body and  the  orientational  diffusion coefficients  $d_\alpha$.
This strategy no  longer requires the solution of  a complex continuum
viscoelastic model in order to  estimate the precession rate, shifting
the  focus  instead to  determining  the  inherent material  constants
\(d_\alpha\).  This may be seen as a drawback, but note that even when
dissipation  and   precession  rates   are  computed   from  continuum
mechanics, one  still faces the  challenge of determining  the unknown
parameters  in  the  rheological   model  being  used.   Our  strategy
distinctively avoids conflating different  levels of description, such
as  combining  Euler's  rigid  body  equations  with  continuum  field
descriptions  of  wobbling  bodies,  and has  a  clear  definition  of
entropy, dissipation, and the different  forms of kinetic energy for a
woobling  body.  As  will  be shown  somewhere  else, the  dissipative
Euler's  equations   can  be   generalized  to  include   an  external
gravitational field. These equations display precession relaxation and
the  phenomenon  of   tidal  locking,  and  both   are  unnafected  by
viscoelasticity.  We  believe that  the dissipative  Euler's equations
(\ref{ODE}) paves the way to an intuitive and effective way to explore
dissipative processes  in celestial  mechanics, closely  mirroring the
elegance and simplicity of Euler's original equations.

PE   appreciates  useful   discussions  with   Michael  Efroimsky.   We
acknowledge the computational resources and assistance provided by the
Centro de  Computación de  Alto Rendimiento CCAR-UNED.   This research
has   been   supported   through   MCIN   grants   PDC2021-121441-C22,
PID2020-117080RB-C54.


\begin{thebibliography}{33}
\expandafter\ifx\csname natexlab\endcsname\relax\def\natexlab#1{#1}\fi
\expandafter\ifx\csname bibnamefont\endcsname\relax
  \def\bibnamefont#1{#1}\fi
\expandafter\ifx\csname bibfnamefont\endcsname\relax
  \def\bibfnamefont#1{#1}\fi
\expandafter\ifx\csname citenamefont\endcsname\relax
  \def\citenamefont#1{#1}\fi
\expandafter\ifx\csname url\endcsname\relax
  \def\url#1{\texttt{#1}}\fi
\expandafter\ifx\csname urlprefix\endcsname\relax\def\urlprefix{URL }\fi
\providecommand{\bibinfo}[2]{#2}
\providecommand{\eprint}[2][]{\url{#2}}

\bibitem[{\citenamefont{Goldstein}(1983)}]{goldsteinClassicalMechanics1983}
\bibinfo{author}{\bibfnamefont{H.}~\bibnamefont{Goldstein}},
  \emph{\bibinfo{title}{Classical {{Mechanics}}}}
  (\bibinfo{publisher}{{Addison-Wesley, Massachusetts}}, \bibinfo{year}{1983}).

\bibitem[{\citenamefont{Poinsot and Whitley}(2022)}]{poinsot2022outlines}
\bibinfo{author}{\bibfnamefont{L.}~\bibnamefont{Poinsot}} \bibnamefont{and}
  \bibinfo{author}{\bibfnamefont{C.}~\bibnamefont{Whitley}},
  \emph{\bibinfo{title}{Outlines of a New Theory of Rotatory Motion}}
  (\bibinfo{publisher}{{Creative Media Partners, LLC}}, \bibinfo{year}{2022}).

\bibitem[{\citenamefont{Landau and
  Lifshitz}(1960)}]{landauMechanicsThirdEdition1960}
\bibinfo{author}{\bibfnamefont{L.~D.} \bibnamefont{Landau}} \bibnamefont{and}
  \bibinfo{author}{\bibfnamefont{E.~M.} \bibnamefont{Lifshitz}},
  \emph{\bibinfo{title}{Mechanics ({{Third Edition}})}}
  (\bibinfo{publisher}{{Pergamon Press}}, \bibinfo{year}{1960}).

  \bibitem[{\citenamefont{Lamy and Burns}(1972)}]{Lamy1972}
\bibinfo{author}{\bibfnamefont{P.~L.} \bibnamefont{Lamy}} \bibnamefont{and}
  \bibinfo{author}{\bibfnamefont{J.~A.} \bibnamefont{Burns}},
  \bibinfo{journal}{American Journal of Physics} \textbf{\bibinfo{volume}{40}},
  \bibinfo{pages}{441} (\bibinfo{year}{1972}).

\bibitem[{\citenamefont{Ashbaugh et~al.}(1991)\citenamefont{Ashbaugh, Chicone,
  and Cushman}}]{Ashbaugh1991}
\bibinfo{author}{\bibfnamefont{M.~S.} \bibnamefont{Ashbaugh}},
  \bibinfo{author}{\bibfnamefont{C.~C.} \bibnamefont{Chicone}},
  \bibnamefont{and} \bibinfo{author}{\bibfnamefont{R.~H.}
  \bibnamefont{Cushman}}, \bibinfo{journal}{Journal of Dynamics and
  Differential Equations} \textbf{\bibinfo{volume}{3}}, \bibinfo{pages}{67}
  (\bibinfo{year}{1991})

\bibitem[{\citenamefont{Veritasium}()}]{veritasium}
\bibinfo{author}{\bibnamefont{Veritasium}}, \emph{\bibinfo{title}{The {{Bizarre
        Behavior}} of {{Rotating Bodies}}}}

\verb|https://www.youtube.com/watch?v=1VPfZ_XzisU|.

\bibitem[{\citenamefont{Unsimplified}()}]{physicsunsimplified}
\bibinfo{author}{\bibfnamefont{P.}~\bibnamefont{Unsimplified}},
\emph{\bibinfo{title}{Rigid {{Body Motion}} and the {{Dzhanibekov Effect}}}}.

\verb|https://www.youtube.com/watch?v=NJLdW4DHRcA|

\bibitem[{\citenamefont{Van~Damme et~al.}(2017)\citenamefont{Van~Damme,
  Marde{\v s}i{\'c}, and Sugny}}]{vandamme2017}
\bibinfo{author}{\bibfnamefont{L.}~\bibnamefont{Van~Damme}},
  \bibinfo{author}{\bibfnamefont{P.}~\bibnamefont{Marde{\v s}i{\'c}}},
  \bibnamefont{and} \bibinfo{author}{\bibfnamefont{D.}~\bibnamefont{Sugny}},
  \bibinfo{journal}{Physica D: Nonlinear Phenomena}
  \textbf{\bibinfo{volume}{338}}, \bibinfo{pages}{17} (\bibinfo{year}{2017}).
  

\bibitem[{\citenamefont{Marde{\v s}i{\'c} et~al.}(2020)\citenamefont{Marde{\v
  s}i{\'c}, Guillen, Van~Damme, and Sugny}}]{mardesic2020a}
\bibinfo{author}{\bibfnamefont{P.}~\bibnamefont{Marde{\v s}i{\'c}}},
  \bibinfo{author}{\bibfnamefont{G.~J.~G.} \bibnamefont{Guillen}},
  \bibinfo{author}{\bibfnamefont{L.}~\bibnamefont{Van~Damme}},
  \bibnamefont{and} \bibinfo{author}{\bibfnamefont{D.}~\bibnamefont{Sugny}},
  \bibinfo{journal}{Physical Review Letters} \textbf{\bibinfo{volume}{125}},
  \bibinfo{pages}{064301} (\bibinfo{year}{2020}).

\bibitem[{\citenamefont{Murakami et~al.}(2016)\citenamefont{Murakami, Rios, and
  Impelluso}}]{Murakami2016}
\bibinfo{author}{\bibfnamefont{H.}~\bibnamefont{Murakami}},
  \bibinfo{author}{\bibfnamefont{O.}~\bibnamefont{Rios}}, \bibnamefont{and}
  \bibinfo{author}{\bibfnamefont{T.~J.} \bibnamefont{Impelluso}},
  \bibinfo{journal}{Journal of Applied Mechanics, Transactions ASME}
  \textbf{\bibinfo{volume}{83}} (\bibinfo{year}{2016}).

\bibitem[{\citenamefont{Trivailo and Kojima}(2019)}]{Trivailo2019}
\bibinfo{author}{\bibfnamefont{P.~M.} \bibnamefont{Trivailo}} \bibnamefont{and}
  \bibinfo{author}{\bibfnamefont{H.}~\bibnamefont{Kojima}},
  \bibinfo{journal}{Transactions of the Japan Society for aeronautical and
  space sciences, Aerospace Technology Japan} \textbf{\bibinfo{volume}{17}},
  \bibinfo{pages}{72} (\bibinfo{year}{2019}).

\bibitem[{\citenamefont{{Lun-Fu} et~al.}(2022)\citenamefont{{Lun-Fu},
  Bubenchikov, Bubenchikov, Kaparulin, and Ovchinnikov}}]{lun-fu2022}
\bibinfo{author}{\bibfnamefont{A.~V.} \bibnamefont{{Lun-Fu}}},
  \bibinfo{author}{\bibfnamefont{A.~M.} \bibnamefont{Bubenchikov}},
  \bibinfo{author}{\bibfnamefont{M.~A.} \bibnamefont{Bubenchikov}},
  \bibinfo{author}{\bibfnamefont{D.~S.} \bibnamefont{Kaparulin}},
  \bibnamefont{and} \bibinfo{author}{\bibfnamefont{V.~A.}
  \bibnamefont{Ovchinnikov}}, \bibinfo{journal}{Meccanica}
  \textbf{\bibinfo{volume}{57}}, \bibinfo{pages}{2293} (\bibinfo{year}{2022})

\bibitem[{\citenamefont{Wheatland et~al.}(2021)\citenamefont{Wheatland, Murphy,
  Naoumenko, Schijndel, and Katsifis}}]{wheatland2021}
\bibinfo{author}{\bibfnamefont{M.~S.} \bibnamefont{Wheatland}},
  \bibinfo{author}{\bibfnamefont{T.}~\bibnamefont{Murphy}},
  \bibinfo{author}{\bibfnamefont{D.}~\bibnamefont{Naoumenko}},
  \bibinfo{author}{\bibfnamefont{D.~V.} \bibnamefont{Schijndel}},
  \bibnamefont{and} \bibinfo{author}{\bibfnamefont{G.}~\bibnamefont{Katsifis}},
  \bibinfo{journal}{American Journal of Physics} \textbf{\bibinfo{volume}{89}},
  \bibinfo{pages}{342} (\bibinfo{year}{2021}).

\bibitem[{\citenamefont{Warner et~al.}(2009)\citenamefont{Warner, Harris, and
  Pravec}}]{Warner2009}
\bibinfo{author}{\bibfnamefont{B.~D.} \bibnamefont{Warner}},
  \bibinfo{author}{\bibfnamefont{A.~W.} \bibnamefont{Harris}},
  \bibnamefont{and} \bibinfo{author}{\bibfnamefont{P.}~\bibnamefont{Pravec}},
  \bibinfo{journal}{Icarus} \textbf{\bibinfo{volume}{202}},
  \bibinfo{pages}{134} (\bibinfo{year}{2009}).
  
\bibitem[{\citenamefont{Kwiecinski}(2020)}]{Kwiecinski2020}
\bibinfo{author}{\bibfnamefont{J.~A.} \bibnamefont{Kwiecinski}},
  \bibinfo{journal}{Monthly Notices of the Royal Astronomical Society}
  \textbf{\bibinfo{volume}{497}}, \bibinfo{pages}{19} (\bibinfo{year}{2020}).

\bibitem[{\citenamefont{Efroimsky}(2001)}]{Efroimsky2001}
\bibinfo{author}{\bibfnamefont{M.}~\bibnamefont{Efroimsky}},
  \bibinfo{journal}{Planetary and Space Science} \textbf{\bibinfo{volume}{49}},
  \bibinfo{pages}{937} (\bibinfo{year}{2001}).

\bibitem[{\citenamefont{{K.H. Prendergast}}(1958)}]{Prendergast1958}
\bibinfo{author}{\bibnamefont{{K.H. Prendergast}}}, \bibinfo{journal}{The
  Astronomical Journal} \textbf{\bibinfo{volume}{63}}, \bibinfo{pages}{412}
  (\bibinfo{year}{1958}).

\bibitem[{\citenamefont{Burns et~al.}(1973)\citenamefont{Burns, Safronov, and
  Gold}}]{Burns1973}
\bibinfo{author}{\bibfnamefont{J.~A.} \bibnamefont{Burns}},
  \bibinfo{author}{\bibfnamefont{V.~S.} \bibnamefont{Safronov}},
  \bibnamefont{and} \bibinfo{author}{\bibfnamefont{T.}~\bibnamefont{Gold}},
  \bibinfo{journal}{Monthly Notices of the Royal Astronomical Society}
  \textbf{\bibinfo{volume}{165}}, \bibinfo{pages}{403} (\bibinfo{year}{1973}).
  

\bibitem[{\citenamefont{Efroimsky and Lazarian}(2000)}]{efroimsky2000}
\bibinfo{author}{\bibfnamefont{M.}~\bibnamefont{Efroimsky}} \bibnamefont{and}
  \bibinfo{author}{\bibfnamefont{A.}~\bibnamefont{Lazarian}},
  \bibinfo{journal}{Monthly Notices of the Royal Astronomical Society}
  \textbf{\bibinfo{volume}{311}}, \bibinfo{pages}{269} (\bibinfo{year}{2000}).

\bibitem[{\citenamefont{Efroimsky et~al.}(2002)\citenamefont{Efroimsky,
  Lazarian, and Sidorenko}}]{Efroimsky2002}
\bibinfo{author}{\bibfnamefont{M.}~\bibnamefont{Efroimsky}},
  \bibinfo{author}{\bibfnamefont{A.}~\bibnamefont{Lazarian}}, \bibnamefont{and}
  \bibinfo{author}{\bibfnamefont{V.}~\bibnamefont{Sidorenko}},
  \bibinfo{journal}{https://arxiv.org/abs/astro-ph/0208489v1}
  p.~\bibinfo{pages}{45} (\bibinfo{year}{2002}), \eprint{astro-ph/0208489v1}.

\bibitem[{\citenamefont{Molina et~al.}(2003)\citenamefont{Molina, Moreno, and
  {Mart{\'i}nez-L{\'o}pez}}}]{Molina2003}
\bibinfo{author}{\bibfnamefont{A.}~\bibnamefont{Molina}},
  \bibinfo{author}{\bibfnamefont{F.}~\bibnamefont{Moreno}}, \bibnamefont{and}
  \bibinfo{author}{\bibfnamefont{F.}~\bibnamefont{{Mart{\'i}nez-L{\'o}pez}}},
  \bibinfo{journal}{Astronomy and Astrophysics} \textbf{\bibinfo{volume}{398}},
  \bibinfo{pages}{809} (\bibinfo{year}{2003}).

\bibitem[{\citenamefont{Sharma et~al.}(2005)\citenamefont{Sharma, Burns, and
  Hui}}]{Sharma2005}
\bibinfo{author}{\bibfnamefont{I.}~\bibnamefont{Sharma}},
  \bibinfo{author}{\bibfnamefont{J.~A.} \bibnamefont{Burns}}, \bibnamefont{and}
  \bibinfo{author}{\bibfnamefont{C.-Y.} \bibnamefont{Hui}},
  \bibinfo{journal}{Monthly Notices of the Royal Astronomical Society}
  \textbf{\bibinfo{volume}{359}}, \bibinfo{pages}{79} (\bibinfo{year}{2005}).

\bibitem[{\citenamefont{Lazarian}(2007)}]{Lazarian2007}
\bibinfo{author}{\bibfnamefont{A.}~\bibnamefont{Lazarian}},
  \bibinfo{journal}{Journal of Quantitative Spectroscopy and Radiative
  Transfer} \textbf{\bibinfo{volume}{106}}, \bibinfo{pages}{225}
  (\bibinfo{year}{2007}).

\bibitem[{\citenamefont{Breiter et~al.}(2012)\citenamefont{Breiter, Rozek, and
  Vokrouhlick{\'y}}}]{Breiter2012}
\bibinfo{author}{\bibfnamefont{S.}~\bibnamefont{Breiter}},
  \bibinfo{author}{\bibfnamefont{A.}~\bibnamefont{Rozek}}, \bibnamefont{and}
  \bibinfo{author}{\bibfnamefont{D.}~\bibnamefont{Vokrouhlick{\'y}}},
  \bibinfo{journal}{Monthly Notices of the Royal Astronomical Society}
  \textbf{\bibinfo{volume}{427}}, \bibinfo{pages}{755} (\bibinfo{year}{2012}).

\bibitem[{\citenamefont{Efroimsky}(2015)}]{Efroimsky2015}
\bibinfo{author}{\bibfnamefont{M.}~\bibnamefont{Efroimsky}},
  \bibinfo{journal}{The Astronomical Journal} \textbf{\bibinfo{volume}{150}},
  \bibinfo{pages}{98} (\bibinfo{year}{2015}).

\bibitem[{\citenamefont{Sharma}(2017)}]{sharmaShapesDynamicsGranular2017}
\bibinfo{author}{\bibfnamefont{I.}~\bibnamefont{Sharma}},
  \emph{\bibinfo{title}{Shapes and {{Dynamics}} of {{Granular Minor Planets The
  Dynamics}} of {{Deformable Bodies Applied}} to {{Granular Objects}} in the
  {{Solar System}}}} (\bibinfo{publisher}{{Springer International Publishing}},
  \bibinfo{year}{2017}).

\bibitem[{\citenamefont{Frouard and Efroimsky}(2018)}]{frouard2018}
\bibinfo{author}{\bibfnamefont{J.}~\bibnamefont{Frouard}} \bibnamefont{and}
  \bibinfo{author}{\bibfnamefont{M.}~\bibnamefont{Efroimsky}},
  \bibinfo{journal}{Monthly Notices of the Royal Astronomical Society}
  \textbf{\bibinfo{volume}{473}}, \bibinfo{pages}{728} (\bibinfo{year}{2018}).


\bibitem[{\citenamefont{de~Groot and Mazur}(1964)}]{deGroot1964}
\bibinfo{author}{\bibfnamefont{S.~R.} \bibnamefont{de~Groot}} \bibnamefont{and}
  \bibinfo{author}{\bibfnamefont{P.}~\bibnamefont{Mazur}},
  \emph{\bibinfo{title}{Non-Equilibrium {{Thermodynamics}}}}
  (\bibinfo{publisher}{{North Holland Publishing Company, Amsterdam}},
  \bibinfo{year}{1964}).

\bibitem[{\citenamefont{Saporta~Katz and Efrati}(2019)}]{SaportaKatz2019}
\bibinfo{author}{\bibfnamefont{O.}~\bibnamefont{Saporta~Katz}}
  \bibnamefont{and} \bibinfo{author}{\bibfnamefont{E.}~\bibnamefont{Efrati}},
  \bibinfo{journal}{Physical Review Letters} \textbf{\bibinfo{volume}{122}},
  \bibinfo{pages}{024102} (\bibinfo{year}{2019}).

\bibitem[{\citenamefont{Peng et~al.}(2021)\citenamefont{Peng, Dai, and
  Niemi}}]{Peng2021}
\bibinfo{author}{\bibfnamefont{X.}~\bibnamefont{Peng}},
  \bibinfo{author}{\bibfnamefont{J.}~\bibnamefont{Dai}}, \bibnamefont{and}
  \bibinfo{author}{\bibfnamefont{A.~J.} \bibnamefont{Niemi}},
  \bibinfo{journal}{New Journal of Physics} \textbf{\bibinfo{volume}{23}}
  (\bibinfo{year}{2021}).

\bibitem[{\citenamefont{Espa{\~n}ol et~al.}(2024)\citenamefont{Espa{\~n}ol,
  Thachuk, and De~La~Torre}}]{espanol2024}
\bibinfo{author}{\bibfnamefont{P.}~\bibnamefont{Espa{\~n}ol}},
  \bibinfo{author}{\bibfnamefont{M.}~\bibnamefont{Thachuk}}, \bibnamefont{and}
  \bibinfo{author}{\bibfnamefont{J.}~\bibnamefont{De~La~Torre}},
  \bibinfo{journal}{European Journal of Mechanics - A/Solids}
  \textbf{\bibinfo{volume}{103}}, \bibinfo{pages}{105184}
  (\bibinfo{year}{2024}).

\bibitem[{\citenamefont{V.~I.~Arnold}(1989)}]{v.i.arnoldMathematicalMethodsClassical1989}
\bibinfo{author}{\bibfnamefont{K.~V.} \bibnamefont{V.~I.~Arnold},
  \bibfnamefont{A.~Weinstein}}, \emph{\bibinfo{title}{Mathematical {{Methods Of
  Classical Mechanics}}}}, Graduate {{Texts}} in {{Mathematics}}
  (\bibinfo{publisher}{{Springer}}, \bibinfo{year}{1989}),
  \bibinfo{edition}{2nd} ed.

\bibitem[{\citenamefont{D{\'i}az}(2019)}]{Diaz2019}
\bibinfo{author}{\bibfnamefont{E.~O.} \bibnamefont{D{\'i}az}},
  \emph{\bibinfo{title}{{{3D Motion}} of {{Rigid Bodies A Foundation}} for
  {{Robot Dynamics Analysis}}}} (\bibinfo{publisher}{{Springer}},
  \bibinfo{year}{2019}).

\end{thebibliography}
\end{document}